\documentclass[preprint,aps,showpacs]{revtex4}
\usepackage{graphicx}
\usepackage{bm} % \bm

\begin{document}
\preprint{hep-ph/0406307}

\title{Studies of resonance conditions on neutrino oscillations in matter}

\author{Yuki Kamo}
 \altaffiliation[Also at ]{Radioisotope Center, Kyushu University, %%
3-1-1 Maidashi Higashi-ku, Fukuoka 812-8582, Japan}
 \email{kamo@sci.kumamoto-u.ac.jp}
\author{Satoshi Yajima}
\author{Yoji Higasida}
\author{Shin-Ichiro Kubota}
\author{Shoshi Tokuo}
\affiliation{%
Department of Physics, Kumamoto University, %%
2-39-1 Kurokami, Kumamoto 860-8555, Japan
}

%\date{\today}
%%%%% abstract %%%%%
\begin{abstract}
We analytically discuss the resonance conditions among several neutrinos
in matter. The discriminant for the characteristic equation of the
Hamiltonian is expressed by the coefficients of the equation. The result
of the computation for the discriminants tells us that the neutrino
energy and the matter density are in inverse proportion to each other
at the resonance states in not only 2- but also 3- and 4-neutrino models. 
\end{abstract}

\pacs{14.60.Pq, 14.60.St, 13.15.+g, 96.40.Tv}

\maketitle
\section{Introduction}
Neutrino oscillations mean transitions among neutrino flavors, and give
an interpretation of neutrino
phenomena\cite{SK98,SNO01,KamLAND03,LSND96} found in the observations of
the solar and the atmospheric neutrinos and the LSND experiment.  
The mass squared differences are some parameters of the neutrino
oscillations, and are determined by some results of the observations and
the experiments mentioned above. In order to explain the results of
those within one framework, the three kinds of mass squared differences
are needed. Existence of them means that four neutrinos have different
masses from each other. It is a reason to discuss the 4-neutrino
oscillation. Four neutrinos are classified into active neutrino flavors 
($\nu_e,\nu_{\mu}, \nu_{\tau}$), which interact with leptons in the weak
interaction, and sterile neutrino ($\nu_s$). The sterile neutrino does
not have the weak interaction. 

Recent analyses of the experiments and the observations disfavor the four
neutrino flavors. The possibility of an oscillation 
$\nu_{e} \leftrightarrow \nu_{s}$ is strongly excluded by the
analyses. However, the result of the LSND experiment causes the maximum
one of the three mass squared differences, and gives grounds for the
4-neutrino models. The upcoming MiniBooNE experiments \cite{BooNE02} 
may form a conclusion about this
discrepancy. Whatever conclusion the experiment leads, it is useful to
consider the neutrino oscillations with the sterile neutrino in the
condition different from that of the experiment.

The neutrino oscillation pattern in vacuum can get modified when the
neutrinos pass through matter. This is known as the
Mikheyev-Smirnov-Wolfenstein (MSW) effects\cite{MSW85}, which can be
described by an effective Hamiltonian. 
The Hamiltonian is expressed by the sum of the vacuum
Hamiltonian and the interaction with the charged and the neutral 
currents\cite{Bilenky99}. In order to write the $4 \times 4$ Hamiltonian
matrix in matter, we introduce the $4 \times 4$ mixing matrix which is an
extension of the $3 \times 3$ Maki-Nakagawa-Sakata-Pontecorvo (MNSP)
matrix\cite{MNS62,Pontecorvo57}.  

Analytical calculations of active 3-neutrino oscillations in matter
have been performed\cite{Ohlsson00,Kamo03,Kimura02,Xing00,Harrison00}.
In our previous paper\cite{Kamo03}, we have given analytic expressions
for the time-evolution operator in the 4-neutrino oscillation and the
transition probabilities in the presence of constant matter densities.

In the results of our calculations, the value of the matter density
which causes the neutrino resonance varies with the average of the
neutrino energies. In general, the resonance occurs when two of the
energy levels approach to each other. However, the calculation to get
the energy levels is difficult because we have to solve the
characteristic equation of the Hamiltonian. Then we propose to use the
discriminant of the characteristic equation made from the Hamiltonian in
matter in order to estimate the neutrino resonance conditions. In this
calculation, we need not solve the characteristic equation.

The outline of the article is as follows.
In Sec.~\ref{chap:formalism},
two kinds of bases to express neutrino states are introduced.
These bases are connected to each other by a mixing matrix.
To describe neutrino oscillations in matter,
the effective Hamiltonian with charged and neutral currents is given.
In Sec.~\ref{chap:ResonanceConditions},
we discuss the resonance conditions of neutrinos due to the
characteristic equation of the effective Hamiltonian in matter. We then
make use of a definition of the discriminant for an algebraic equation
and a relation of the coefficients and solutions for the equations. 
In Sec.~\ref{chap:NumericalAnalyses},
the discriminant is concretely computed within a range expected that
the neutrino mixings occur. And the graphical expressions of the
discriminant are illustrated.
Finally, we discuss the resonance conditions for the neutrino mixing in
matter in Sec.~\ref{chap:Discussion}.

\section{Formalism}\label{chap:formalism}
We analyze the matter effects of 4-neutrino oscillation%
\cite{Kamo03}, similar to those of 2- and 3-neutrino oscillations. 

\subsection{Two bases and a mixing matrix}
Neutrinos are produced in flavor eigenstates
$ |{\nu_{\alpha}}\rangle (\alpha = e, \mu, \tau, s)$.
Between the source and the detector, the neutrinos evolve as 
mass eigenstates 
$ |{\nu_{a}}\rangle (a = 1, 2, 3, 4)$.
There are two kinds of eigenstates:
$ |{\nu_{\alpha}}\rangle$ and $|{\nu_{a}}\rangle$.
These eigenstates are defined by neutrino fields
$\nu_{\alpha}$ and $\nu_{a}$ corresponding to each eigenstate:
$\nu^{\dagger} |{0}\rangle \equiv |{\nu}\rangle$,
$|{\nu_{\alpha}}\rangle \equiv |{\alpha}\rangle$,
$|{\nu_{a}}\rangle \equiv |{a}\rangle$,
where a vacuum state is given by $|{0}\rangle$.
In the present analysis, 
we will use the plane wave approximation of the fields.
In this approximation, a neutrino flavor field $\nu_{\alpha}$ is
expressed by a linear combination of neutrino mass field $\nu_{a}$:
\begin{equation}
 \nu_{\alpha} = \sum _{a=1}^{4} U_{\alpha a} \nu_{a},
  \label{Eq:plane-wave-1}
\end{equation}
where $U$ is a 4 $\times$ 4 unitary matrix 
with the elements $U_{\alpha a}$.
If we write this relation in neutrino eigenstates, then
\begin{equation}
 |{\alpha}\rangle = \sum_{a=1}^{4} U_{\alpha a}^{*} |{a}\rangle.
\end{equation}

An arbitrary neutrino state $\psi$ is expressed in both the flavor and
the mass bases as 
\begin{eqnarray}
 \psi &\equiv& \sum_{\alpha = e, \mu, \tau, s} \psi_{\alpha} |{\alpha}\rangle
  =
  \sum_{a=1}^{4} \psi_{a}|{a}\rangle,
\end{eqnarray}
where $\psi_{\alpha}$ and $\psi_{a}$ are 
the components of $\psi$ of the flavor eigenstate basis and
the mass eigenstate basis, respectively.
They are related to each other in the form
\begin{equation}
 \psi_{a} = \sum_{\alpha = e, \mu, \tau, s} U_{\alpha a}^{*} \psi_{\alpha}.
  \label{eq:relation-psi_a-psi_alpha}
\end{equation}
If we define the matrix elements as
\begin{eqnarray}
 \psi^{\mathrm{(flavor)}} &=& (\psi_{\alpha}) 
  = \left( 
     \begin{array}{c} 
      \psi_e \\
      \psi_{\mu} \\
      \psi_{\tau} \\
      \psi_{s}
     \end{array}
   \right),
\quad
 \psi^{\mathrm{(mass)}} = (\psi_{a}) 
 = \left( 
    \begin{array}{c}
     \psi_1 \\
     \psi_2 \\
     \psi_3 \\
     \psi_4
    \end{array}
  \right),
 \\
 U &=& (U_{\alpha a}) =
  \left(
   \begin{array}{cccc}
    U_{e1} & U_{e2} & U_{e3} & U_{e4} \\
    U_{\mu 1} & U_{\mu 2} & U_{\mu 3} & U_{\mu 4} \\
    U_{\tau 1} & U_{\tau 2} & U_{\tau 3} & U_{\tau 4} \\
    U_{s1} & U_{s2} & U_{s3} & U_{s4}
   \end{array}
 \right),
  \label{eq:unitary-matrix}
\end{eqnarray}
the relations (\ref{eq:relation-psi_a-psi_alpha}) of the components
between the flavor and the mass eigenstates are rewritten in the matrix
form, 
\begin{equation}
 \psi^{\mathrm{(flavor)}} = U \psi^{\mathrm{(mass)}}.
\end{equation}
The unitary matrix $U$ is the mixing matrix of four neutrinos. 
There are 6 mixing angles and 3 phases
as parameters of $U$, in the case of 4 neutrinos.

A parameterization for $U$ is given by
\begin{equation}
 U = R_{34} \tilde{R}_{24} \tilde{R}_{14} R_{23} \tilde{R}_{13} R_{12},
  \label{eq:DefMixingMatrix}
\end{equation}
where $R_{ab}$ and $\tilde{R}_{ab}$ are the real and the complex mixing
matrices, respectively\cite{Fritzsch87}. The structure of these matrices is 
\begin{eqnarray}
 {(R_{ab})}_{ij} &=& 
  \delta_{ij} + (\cos \theta_{ab} - 1)(\delta_{ai}\delta_{aj} 
 + \delta_{bi}\delta_{bj})
 \nonumber \\ 
 && +
 {\sin \theta_{ab}} 
 (\delta_{ai}\delta_{bj} - \delta_{bi}\delta_{aj})
 \label{eq:DefRealR}
 , \\ %%%%%
 {(\tilde{R}_{ab})}_{ij} &=& \delta_{ij} 
 + ({\cos \theta_{ab}} - 1)
 (\delta_{ai}\delta_{aj} + \delta_{bi}\delta_{bj})
 \nonumber \\
 && +
 {\sin \theta_{ab}} 
 ({e^{-i \Delta_{ab}}}\delta_{ai}\delta_{bj}
 -
 {e^{i \Delta_{ab}}}\delta_{bi}\delta_{aj}),
 \label{eq:DefComplexR}
\end{eqnarray}
where the mixing of mass eigenstates $|a \rangle$ and $|b\rangle$ is
described by 6 mixing
angles $\theta_{ab}$: $\theta_{12}$, $\theta_{13}$, $\theta_{14}$,
$\theta_{23}$, $\theta_{24}$, $\theta_{34}$ and 3 phases $\Delta_{ab}$: 
$\Delta_{13}$, $\Delta_{24}$, $\Delta_{34}$. In the case of leaving the
CP violations out of consideration, i.e., $\Delta_{ab}=0$, 
$\tilde{R}_{ab} = R_{ab}$, and the mixing
matrix $U$ is a real orthogonal matrix\cite{Kamo03}.

\subsection{Hamiltonian in matter}
In the mass eigenstate basis, the Hamiltonian ${\mathcal{H}}_0$ 
participating in the propagation of neutrinos in vacuum is given by
\begin{equation}
 {\mathcal{H}}_0 = \mathrm{diag}({E_1, E_2, E_3, E_4}),
\end{equation}
where $E_a ~ (a=1,2,3,4)$ are the energies of 
the neutrino mass eigenstates $|{a}\rangle$ with mass 
$m_a$:
\begin{equation}
 E_a = \sqrt{{m_a}^2 + \bm{p}^2}.\label{Eq:on-shell}
\end{equation}
Here and hereafter,
we assume the momentum $\bm{p}$ to be the same for all mass eigenstates.
If we assume $|\bm{p}| \gg m_a$, 
\begin{equation}
  E_a \sim |\bm{p}| + \frac{{m_a}^2}{2|\bm{p}|} 
   \sim |\bm{p}| + \frac{{m_a}^2}{2E},
   \label{eq:approximationEa}
\end{equation}
where $E$ is the average of the neutrino energies.

The interactions between neutrinos and matter are described by two kinds of
additional potentials. One is the interaction of the charged particles
(electrons) and its neutrino $\nu_e$:
\begin{equation}
 V_{\mathrm{CC}} = V_e = \sqrt{2} G_F ~\mathrm{diag}(N_e, 0, 0, 0).
\end{equation}
The other is the interaction of the neutral particles(e.g., the neutron)
and active neutrinos ($\nu_e,\nu_{\mu},\nu_{\tau}$):
\begin{equation}
 V_{\mathrm{NC}} = \sqrt{2} G_F ~
  \mathrm{diag}(-\frac{1}{2}N_n, -\frac{1}{2}N_n, -\frac{1}{2}N_n, 0),%
  \label{Eq:NC-interacton}
\end{equation}
where $G_F$, $N_e$, and $N_n$ are the Fermi weak coupling constant, the
electron number density and the neutral particle number density,
respectively. Note that we assume the particle number densities to be
constant throughout the matter where the neutrinos are propagating. The
interaction term (\ref{Eq:NC-interacton}) can be separated into two
parts as
\begin{eqnarray}
 V_{\mathrm{NC}} &=& V_n + V^{\prime}, \\
 V_n &=& \sqrt{2} G_F ~ \mathrm{diag}(0, 0, 0, +\frac{1}{2}N_n),\\
 V^{\prime} &=& \sqrt{2} G_F ~
  \mathrm{diag}(-\frac{1}{2}N_n, -\frac{1}{2}N_n, -\frac{1}{2}N_n, -\frac{1}{
2}N_n).
\end{eqnarray}

The interaction terms $V_e$ and $V_n$ are added to vacuum Hamiltonian
$\mathcal{H}_0$ for the propagation of neutrinos in matter. However, the
vacuum Hamiltonian and the interaction terms are written by the mass and
the flavor eigenstates, respectively. 

The mass eigenstate is convenient for describing the neutrinos
propagation in matter. The Hamiltonian in the mass basis is 
\begin{eqnarray}
 {\mathcal{H}}_m^{\mathrm{(mass)}} 
  &=& 
  {\mathcal{H}}_0 + U^{-1}V_{\mathrm{CC}} U + U^{-1}V_{\mathrm{NC}} U
  \nonumber \\
 &=&
  {\mathcal{H}}_0 + U^{-1}V_{e} U + U^{-1}V_{n} U -A_n I
  \label{eq:HamiltonianInMatter},
\end{eqnarray}
where $I$ is the $4 \times 4$ unit matrix. The matter densities $A_e$,
$A_n$ and $A$ are defined by
\begin{eqnarray}
  A_e &=& \sqrt{2} G_F N_e \equiv A, \label{eq:defA-e}\\
 A_n &=& \frac{1}{\sqrt{2}} G_F N_n = 
  \frac{1}{2} A \frac{N_n}{N_e}. \label{eq:defA-n}
\end{eqnarray}
However, the interaction terms $U^{-1} V_e U$ and $U^{-1} V_n U$ are
complicated\cite{Kamo03}.

The flavor eigenstate is convenient for describing the interaction. The
Hamiltonian in the flavor basis is 
\begin{eqnarray}
 {\mathcal{H}}_m^{(\mathrm{flavor})} 
  &=& 
  {U {\mathcal{H}}_0 U^{-1}} + V_{\mathrm{CC}}+ V_{\mathrm{NC}}
  \nonumber \\
 &=&
  {U {\mathcal{H}}_0 U^{-1}} + V_e+ V_n - A_n I \nonumber \\
 &=& {U {\mathcal{H}}_0 U^{-1}} +
  \mathrm{diag}(A_e, 0, 0, A_n) -A_n I
  \label{eq:HamiltonianInFlavor}.
\end{eqnarray}
Although the term of a vacuum Hamiltonian is complicated,
the interaction terms are simple for matter densities $A_e$ and $A_n$.

\section{Calculations of resonance conditions}
\label{chap:ResonanceConditions}
Some analytical calculations show that the transition probabilities
for the neutrino oscillations in matter depend on the matter density%
\cite{Ohlsson00,Kamo03}. In these analyses, the neutrino resonance
occurs at the critical (matter) density. The critical density also
changes with variation of neutrino energy. The lower neutrino energy
reduces, the higher the density to observe the resonance of neutrino becomes. 

The resonance of neutrino occurs when the energy levels in the presence of
matter approach to each other. To discuss the resonance conditions, we
consider the relations between the neutrino energy and the matter
density. The neutrino energy is one of eigenvalues of the characteristic
equation for the matter Hamiltonian. Therefore, the behavior of the
eigenvalues decides the neutrino resonance conditions.

\subsection{Discriminant of the characteristic equation}
To simplify discussion, we consider the Hamiltonian
$\mathcal{H}_m^{\mathrm{(flavor)}}$ in the flavor basis in 
(\ref{eq:HamiltonianInFlavor}), in which the interaction terms are
simple for matter density $A$. The degree of the characteristic equation
of the Hamiltonian is determined by 
the number of neutrino flavors we suppose, e.g., the
characteristic equation for 3-neutrino oscillations is a cubic
one. The roots $\alpha_{i}$ ($i=1,2,3,\cdots$) of the characteristic
equation become more complicated as the degree of the equation
increases. It is tedious to solve the characteristic equation of higher
degree. 

In general, the discriminant $D_{(n)}$ of the characteristic equation
\begin{equation}
 x^n + c_{n-1}x^{n-1} + c_{n-2}x^{n-2} + \cdots + c_0 = 0
  \label{eq:characteristic-eq-n}
\end{equation}
with degree $n$, which has constant coefficients
$c_{k}$ $(k = 0,1,\cdots, n-1)$, are expressed by the roots $\alpha_{i}$
with respect to $x$ of the equation:
\begin{equation}
 D_{(n)} = \prod_{1 \leq i < j \leq n} {(\alpha_i - \alpha_j)}^2.
  \label{eq:discriminant-def}
\end{equation}
The discriminant $D_{(n)}$ includes the difference between eigenvalues 
$\alpha_i$ and $\alpha_j$, that is, a effective energy difference. It
suggests that we can use the discriminant $D_{(n)}$ to judge the
neutrino resonance conditions. If $D_{(n)}$ becomes small, we can
conclude that the resonance occurs, because $D_{(n)}$ depends on the
square ${(\alpha_i - \alpha_j)}^2$ of any energy differences. Therefore,
some resonances may occur when one of ${(\alpha_i - \alpha_j)}^2$ has
(local) minimum value. 

The discriminant $D_{(n)}$ is polynomials of the roots $\alpha_i$, of
which degree is $2\times\,_nC_2$, e.g., 6 for the discriminant of the
cubic equation. The discriminant $D_{(n)}$ is also written as the
symmetric polynomials for $\alpha_i$. The fundamental
theorem of symmetric polynomials tells us that every symmetric
polynomials in $\alpha_{i}$ ($i=1,2,3,\cdots,n$) can be represented by
the following elementally symmetric polynomials: 
\begin{eqnarray}
 \alpha_1 + \alpha_2 + \cdots + \alpha_n
 &=& c_{n-1},
  \nonumber\\
 \alpha_1\alpha_2 + \alpha_1\alpha_3 + \cdots + \alpha_{n-1}\alpha_n
  &=& c_{n-2}, 
  \nonumber\\ 
 \alpha_1\alpha_2\alpha_3 + \cdots + \alpha_{n-2}\alpha_{n-1}\alpha_n
  &=& c_{n-3}, 
 \nonumber\\
 \cdots &&, \nonumber\\
 \alpha_1\alpha_2\cdots\alpha_n &=& c_0.
  \label{Eq:CoefRootPolynomial}
\end{eqnarray}
The solutions $\alpha_i$ and the coefficients $c_k$ for the
characteristic equation are related to each other. Indeed, this
relations between $\alpha_i$ and $c_k$ are given by
(\ref{Eq:CoefRootPolynomial}). 
In brief, the discriminant $D_{(n)}$, which includes the effective
neutrino energy differences, can be represented by the coefficients of
the characteristic equation for the Hamiltonian in matter without
solving the characteristic equation. 

For example, a discriminant $D_{(2)}$ for a algebraic equation
$x^2 + c_0 =0$ is calculated as follows:
\begin{eqnarray}
 D_{(2)} \equiv {(\alpha_1 - \alpha_2)}^2 
  &=&
 {(\alpha_1 + \alpha_2)}^2 - 4 \alpha_1 \alpha_2
 \nonumber \\
 &=&
  -4 c_0.
  \label{eq:discriminant2}
\end{eqnarray}
The concrete form of discriminants $D_{(3)}$ and $D_{(4)}$ are given as
follows: 
\begin{eqnarray}
 D_{(3)} &=& -27 {c_0}^2 - 4{c_1}^3 \quad
  (\mathrm{for} \  x^3 + c_1 x + c_0 = 0), 
    \label{eq:discriminant3}\\
 D_{(4)} &=& 
  256 {c_0}^3 -27 {c_1}^4 +144 c_0 {c_1}^2 c_2 \nonumber \\
 && -128 {c_0}^2{c_2}^2  -4 {c_1}^2 {c_2}^3 +16 c_0 {c_2}^4 \nonumber \\
 && (\mathrm{for} \ x^4 + c_2 x^2 +c_1 x + c_0 = 0)
  \label{eq:discriminant4}.
\end{eqnarray}
Note that, for the algebraic equation
(\ref{eq:characteristic-eq-n}) with degree $n$, a term $c_{n-1}x^{n-1}$
with power $(n-1)$  of a variable $x$ vanishes by a suitable
transformation: $x \to x - c_{n-1}/n$.

\subsection{Traceless Matrix $T_N$}
In order to find the explicit form of the characteristic equation, the
Hamiltonian in the matrix form is separated into the diagonal and the
traceless matrices. 
An arbitrary $N \times N$ matrix $M$ can always be written as
\begin{equation}
 M = T_N + \frac{1}{N} ({\mathrm{tr}} M) I_N
  \label{eq:M-2-M01},
\end{equation}
where $T_N$ and $I_N$ are 
$N \times N$ traceless and unit matrices, respectively.
Note that $\mathrm{tr} T_N = 0 $.

The characteristic equation for the matrix $T_N$ is written by an eigenvalue
$\lambda$ of $T_N$ as
\begin{eqnarray}
 0 &=&
 \mathrm{det}(T_N- \lambda I_N) \nonumber \\
 &=& 
 {\lambda}^N + c_{N-1} {\lambda}^{N-1} + \cdots
 + c_2 {\lambda}^2 + c_1 {\lambda}+ c_0 
  \label{eq:N-eigen-eq},
\end{eqnarray}
where the coefficient $c_0$ is the determinant of $T_N$, 
e.g., $c_0= \mathrm{det} T_N$ and 
$c_k$ \ ($ k= 1,2 \cdots N-1$) are expressed by the sum of cofactors of the
diagonal elements of $T_N$. But $c_{N-1}=0$, because $c_{N-1}$ is equal
to the trace of the traceless matrix $T_N$. 

In the 4-neutrino model, the Hamiltonian in the matrix form is given in 
(\ref{eq:HamiltonianInMatter}) and (\ref{eq:HamiltonianInFlavor}).
Ignoring the CP violations, the trace of
the matrix 
${\mathcal H}_m^{\mathrm{(flavor)}}$ in (\ref{eq:HamiltonianInFlavor}) is
\begin{equation}
 {\mathrm{tr}} \mathcal{H}_m^{\mathrm{(flavor)}} 
  = E_1 + E_2 + E_3 + E_4 + A_e -3 A_n,
\end{equation}
where we use the unitary conditions, e.g.,
${U_{e1}}^2 +{U_{e2}}^2 +{U_{e3}}^2 +{U_{e4}}^2 = 1$.
Then, the Hamiltonian in matter is described as
\begin{equation}
 \mathcal{H}_m^{\mathrm{(flavor)}}= 
  T_4 + \frac{1}{4}({\mathrm{tr}} {\mathcal{H}}_m^{\mathrm{(flavor)}}) I_4.
\end{equation}

The characteristic equation for matrix $T$ is written as 
\begin{equation}
 {\lambda}^4 + c_3 {\lambda}^3 + c_2 {\lambda}^2 + c_1 {\lambda}+ c_0 
 = 0
  \label{eq:4d-eigen-eq}.
\end{equation}
The coefficients $c_l$ ($l=1,2,3,4$) are expressed in the matrix form as 
\begin{eqnarray*}
 c_0 &=& \textrm{det}\,T,\\
 c_1 &=& - \textrm{cof}\,T_{(1)} - \textrm{cof}T_{(2)}
  - \textrm{cof}\,T_{(3)} - \textrm{cof}T_{(4)}, \\
 c_2 &=&   \textrm{cof}\,T_{(12)} + \textrm{cof}\,T_{(13)}
  + \textrm{cof}\,T_{(14)} \nonumber \\ &&
 + \textrm{cof}\,T_{(23)}
  + \textrm{cof}\,T_{(24)} + \textrm{cof}\,T_{(34)},\\
 c_3 &=& - {\mathrm{tr}} T = 0,
\end{eqnarray*}
where the cofactors $\textrm{cof}\,T_{(p)}$ $(p=1, 2, 3, 4)$ of 
diagonal components $T_{pp}$ of $T_4$
and $\textrm{cof}\,T_{(rs)}$ $(1 \leq r < s \leq 4)$ of
$T_{rr}$ and $T_{ss}$ are determinants of $3 \times 3$ and 
$2 \times2$ matrices, respectively, e.g. ,
\begin{eqnarray*}
 \textrm{cof}\,T_{(2)} &=&
  \sum_{p_1, p_2, p_3, p_4=1}^4 \epsilon_{p_1 p_2 p_3 p_4}
  T_{1 p_1}\delta_{2 p_2} T_{3 p_3} T_{4 p_4}, \\
 \textrm{cof}\,T_{(13)} &=& 
  \sum_{p_1, p_2, p_3, p_4=1}^4 \epsilon_{p_1 p_2 p_3 p_4}
  \delta_{1 p_1} T_{2 p_2} \delta_{3 p_3} T_{4 p_4}.
\end{eqnarray*}

The coefficients $c_{k}$ of the characteristic equation for the matrix
$T_4$ are obtained from the components $T_{rs}$ of $T_4$, and the
discriminant $D_{(4)}$ is derived by these coefficients from 
(\ref{eq:discriminant4}). The discriminants of $T_2$ and $T_3$ in 2- and
3-neutrino models, respectively, are given in the similar method as in
4-neutrino model.

\subsection{Resonance conditions for 2-neutrino oscillation}
The discriminant of the characteristic equation for the matter
Hamiltonian is derived by the calculations shown below. For the simple
case, a calculation for the discriminant of 2-neutrino Hamiltonian is
performed.  

The matter Hamiltonian for $\nu_{\alpha}$ and $\nu_{\beta}$ is described
as 
\begin{eqnarray}
{\mathcal H}_m^{(\mathrm{flavor})}
 &=& U {\mathcal H}_0^{\mathrm{(mass)}} U^{-1} 
 + {V_{\mathrm{CC}}} 
 + {V_{\mathrm{NC}}}  
 \nonumber \\ 
 &=&
  \left[
   \begin{array}{ll}
    E - \frac{E_{21}}{2} \cos 2\theta 
     &
     \frac{E_{21}}{2} \sin 2\theta \\
    \frac{E_{21}}{2} \sin 2\theta &
     E + \frac{E_{21}}{2} \cos 2\theta 
   \end{array}
  \right]
  \nonumber \\
  &&+
  \left[
   \begin{array}{ll}
    {A_e \delta_{e\alpha}}
     &
     0
     \\
    0
     &
     {A_n \delta_{s\beta}}
   \end{array}
	      \right],
  \label{eq:matterHamiltonianfor2nu}
\end{eqnarray}
where $\theta$ is the mixing angle, and
\begin{equation}
 E_{21}=E_2 - E_1,\quad E=\frac{1}{2}(E_1 + E_2).
\end{equation}
The interaction term in (\ref{eq:matterHamiltonianfor2nu})
includes the factors $A_e \delta_{e\alpha}$ and $A_n \delta_{s\beta}$.
These factors appear if we assume the electron neutrino $\nu_e$ and the
sterile neutrino $\nu_s$, respectively. In the case of 2-neutrino
Hamiltonian for $\nu_e$ and $\nu_{\mu}$, we can leave $A_n \delta_{s\beta}$ 
out of consideration. 

From (\ref{eq:M-2-M01}), the Hamiltonian (\ref{eq:matterHamiltonianfor2nu}) is rewritten
for the $2 \times 2$ traceless matrix $T_2$:
%\begin{equation}
\begin{eqnarray}
 {\mathcal H}_m^{\mathrm{(flavor)}} 
  &=&
  T_2 + \frac{1}{2}(\mathrm{tr} {\mathcal H}_m^{\mathrm{(flavor)}})I_2,
  \\
 \mathrm{tr} {\mathcal H}_m^{\mathrm{(flavor)}} 
  &=& 2E + A_e \delta_{e\alpha}+ A_n \delta_{s\beta}.
\end{eqnarray}
The concrete form of the traceless matrix $T_2$ is
\begin{eqnarray}
 T_2 &=& \frac{E_{21}}{2}
  \left[
      \begin{array}{rr}
    - \cos 2\theta & \sin 2\theta \\
       \sin 2\theta & \cos 2\theta
      \end{array}
     \right]
  \nonumber \\
     &&+ \frac{1}{2}(A_e\delta_{e\alpha} -A_n\delta_{s\beta})
     \mathrm{diag}(1,-1).
\end{eqnarray}

The characteristic equation for the matrix $T_2$ is expressed as
\begin{eqnarray}
 0 &=& \mathrm{det}(T_2-\lambda I_2)
  \nonumber \\
 &=&
  {\lambda}^2 -\frac{1}{4} 
  \left[
   ({A_e}\delta_{e\alpha} 
   - {A_n}\delta_{s\beta}
   - {E_{21}} \cos 2\theta )^2 
   + {{E_{21}}}^2 \sin ^2 2\theta
  \right].
  \nonumber
\end{eqnarray}
Then, the discriminant $D_{(2)}$ is presented as follows
\begin{equation}
D_{(2)} = (A_e\delta_{e\alpha} -{A_n}\delta_{s\beta}
       -{E_{21}} \cos 2\theta )^2 
       +E_{21}^2 \sin ^2 2\theta.
       \label{eq:Discriminant-2Nu}
\end{equation}
Therefore, the condition for the neutrino resonance, where
(\ref{eq:Discriminant-2Nu}) gets (local) minimum value, for two neutrinos
$\nu_{\alpha}$ and $\nu_{\beta}$ is  
\begin{equation}
 A_e \delta_{e\alpha} - A_n \delta_{s\beta}
 = 
 E_{21} \cos 2\theta.
 \label{eq:ResonanceConditon2Nu}
\end{equation}
Note that energy differences $E_{21}$ are approximately given by 
\begin{equation}
 E_{21} = E_2 - E_1
  \sim \frac{{m_2}^2 - {m_1}^2}{2E}
  \equiv \frac{\Delta m_{21}^2}{2E},
  \label{eq:differrenceEab}
\end{equation}
where (\ref{eq:approximationEa}) has been used, and $\Delta m_{21}^2$ is a
squared mass difference.

From (\ref{eq:ResonanceConditon2Nu}) and
(\ref{eq:differrenceEab}), we conclude that, for the 2-neutrino
resonance conditions, the average $E$ of the neutrino energies and the
factor ($A_e \delta_{e\alpha} - A_n \delta_{s\beta}$) of the
matter densities are in inverse proportion to each other:
\begin{equation}
 E (A_e \delta_{e\alpha} - A_n \delta_{s\beta})
  =
  \frac{\Delta m_{21}^2}{2}\cos 2\theta.
  \label{eq:ResonanceConditon2NuGeneral}
\end{equation}
Note that (\ref{eq:ResonanceConditon2NuGeneral}) gives a usual
expression\cite{Haxton01} for the resonance condition if we set 
$A_n \delta_{s\beta}=0$. 

\section{Numerical analyses of resonance conditions}
\label{chap:NumericalAnalyses}
We apply some results obtained in the previous section to the 2-, 3- and
4-neutrino models. A discriminant explicitly contains the neutrino energy
differences  $E_{ab} = E_a - E_b$ ($a,b = 1,2,3,4$ in the 4-neutrino
model) in vacuum. These energy differences are approximately given
by the matter mass differences $\Delta m_{ab}^2 = {m_a}^2 - {m_b}^2$,
which are well-known quantities in the various neutrino
oscillation experiments. In the similar way as showed in
(\ref{eq:differrenceEab}), 
\begin{equation}
 E_{ab} \simeq \frac{\Delta {m_{ab}}^2}{2E},
\end{equation}
where $E$ is a average of the neutrino energies, i.e., in four-neutrino
model, $E=(E_1+E_2+E_3+E_4)/4$. 

\begin{figure}[htb]
 \begin{center}
  \includegraphics[height=5cm,clip]{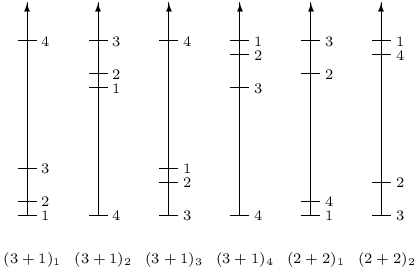}
 \end{center}
 \caption{Several mass patterns for 4-neutrino schemes.}
 \label{fig:massscheme}
\end{figure}

These days, three kinds of mass squared differences are known. They are
represented by $\Delta m_{\mathrm{solar}}^2$, 
$\Delta m_{\mathrm{atm}}^2$ and $\Delta m_{\mathrm{LSND}}^2$, which are 
used as the parameters in the solar and atmospheric oscillations and the
LSND experiment. Using these mass squared differences, several distinct
types of mass patterns are possible\cite{Bilenky99}. The phenomenology and the
mixing matrix depend on the type of the mass scheme. We concentrate on
the discussion on the $(3+1)_1$ - scheme\cite{Kamo03}, which is one of
the several mass patterns in Fig. \ref{fig:massscheme}. This mass-scheme
includes the usual 3-neutrino mass pattern.

In the mass pattern $(3+1)_1$, there are three close masses and
one distinct mass. Let $m_4$ and $\Delta m_{43}^2$ be the distinct mass
and the largest mass squared difference, respectively. Three kinds of the
neutrino mass squared differences are put as follows \cite{PDG02}:
\begin{eqnarray}
 \Delta m_{21}^2 &=& \Delta m^2_{\mathrm{solar}}
  \simeq 7 \times 10^{-5} \,\mathrm{eV^2}, 
  \label{eq:DeltaMsolar} \\
 \Delta m_{32}^2 &=& \Delta m^2_{\mathrm{atm}} 
  \simeq 3 \times 10^{-3} \,\mathrm{eV^2}, 
  \label{eq:DeltaMatm} \\
 \Delta m_{41}^2 &=& \Delta m^2_{\mathrm{LSND}}
  \simeq 1 \,\mathrm{eV^2}
  \label{eq:DeltaMlsnd}.
\end{eqnarray}
Then, the energy differences $E_{ab}$ are expressed by
\begin{eqnarray}
 E_{21} \simeq \frac{\Delta m_{21}^2}{2E},\quad
 E_{32} \simeq \frac{\Delta m_{32}^2}{2E},\quad
 \nonumber \\
 E_{41} \simeq \frac{\Delta m_{41}^2}{2E},\quad
% \nonumber \\
 E_{31} = E_{32} + E_{21},\quad
 \nonumber \\
 E_{42} = E_{41} - E_{21},\quad
 E_{43} = E_{41} - E_{31},\quad
\end{eqnarray}
where we suppose that the average $E$ of the neutrino energies 
is $10 \,\mathrm{GeV}$ or $10 \,\mathrm{MeV}$.

%%%%
Next, we consider the approximate mixing matrix for the $(3+1)_1$-scheme
\cite{Barger00}:
\begin{equation}
\left(
\begin{array}{cccc}
 \frac{1}{\sqrt{2}}{\cos {\epsilon} \, \cos \delta} \, & \,
  {\frac{1}{\sqrt{2}}{\cos \epsilon \, \cos \delta}} \, & \,
  {\cos{\epsilon}\,\sin{\delta}} \, & \,
  {\sin{\epsilon}} 
  \\
 {-\frac{1}{2}} - \frac{1}{2}\sin \delta &
  {\frac{1}{2}} - \frac{1}{2}{\sin \delta} &
  {\frac{1}{\sqrt{2}}}{\cos \delta} &
  0 
  \\
 {\frac{1}{2}} - \frac{1}{2}{\sin \delta} &
  {-\frac{1}{2}} - \frac{1}{2}{\sin \delta}&
  {\frac{1}{\sqrt{2}}}{\cos \delta} &
  0\\
 {\frac{-1}{\sqrt{2}}}{\sin \epsilon} \cos \delta &
  \frac{-1}{\sqrt{2}}{\sin \epsilon} \cos \delta &
  -{\sin \delta \sin \epsilon} &
  \cos {\epsilon}
\end{array} 
\right),
 \label{Eq:31mixingmatrix}
\end{equation}
where $\delta=\frac{5}{180}\pi$ and $\epsilon$ are small: 
$0 \leq \epsilon \leq 0.1 $.
Here the $3 \times 3$ sub-matrix that describes the mixing of 
the three active neutrinos has the bi-maximal form.
The mixing matrix of (\ref{Eq:31mixingmatrix}) 
is given from (\ref{eq:DefMixingMatrix}), (\ref{eq:DefRealR}) and
(\ref{eq:DefComplexR})  by taking 
\begin{eqnarray}
 \theta_{12} = \frac{\pi}{4},\quad
  \theta_{23} = \frac{\pi}{4},\quad
  \theta_{13} = \delta \nonumber \\
  \theta_{14} = \epsilon,
  \quad \theta_{24} = 0 ,\quad \theta_{34} = 0,
  \nonumber \\
  \Delta_{13} = \Delta_{14} = \Delta_{24} = 0.
  \label{eq:MixingAngles}
\end{eqnarray}

In 2- and 3-neutrino models, we can also discuss the resonance condition by a
sub-matrix of $4 \times 4$ mixing matrix. For example, in order to consider the
3-neutrino model, we may ignore the fourth elements in
(\ref{Eq:31mixingmatrix}),  taking $\epsilon=0$. In this case,
$4 \times 4$ mixing matrix (\ref{Eq:31mixingmatrix}) is handled as $3
\times 3$ matrix and a unity element.

In this setting, we can get a numerical expression of the discriminant
$D_{(3)}$ and $D_{(4)}$ as a smooth function of $A$:
\begin{eqnarray}
 D_{(3)} 
  \sim \frac{1}{E^6}
 &&
  \big[
   6.5 \times 10^{-21} -8.4 \times 10^{-18} (AE) 
  \nonumber \\ 
 &&
  + 5.2 \times 10^{-12}{(AE)}^2 - 6.8 \times 10^{-9} {(AE)}^3 \nonumber \\
 &&
 + 2.3 \times 10^{-6} {(AE)}^4
  \big],
  \label{eq:D3num}
\\
 D_{(4)} 
  \sim \frac{1}{E^{12}}
 &&
  \big[
   4.2 \times 10^{-22} -1.3 \times 10^{-19} (AE) 
  \nonumber \\ 
 &&
  + 8.1 \times 10^{-14}{(AE)}^2 - 1.0 \times 10^{-10} {(AE)}^3 \nonumber \\
 &&
  + 3.5 \times 10^{-8}{(AE)}^4 + 7.1 \times 10^{-8} {(AE)}^5 \nonumber \\
 &&
  - 3.2 \times 10^{-8}{(AE)}^6 - 1.4 \times 10^{-7} {(AE)}^7 \nonumber \\
 &&
  - 3.4 \times 10^{-8}{(AE)}^8 + 7.1 \times 10^{-8} {(AE)}^9 \nonumber \\
 &&
  + 3.5 \times 10^{-8}{(AE)}^{10} \nonumber \\
 &&
 + 1.7 \times 10^{-18} {(AE)}^{11}
  \big],
  \label{eq:D4num}
\end{eqnarray} 
where $\epsilon=0.1$.
The derivative $\frac{d D_{(n)}}{d A}$ of $D_{(n)}$ may be useful to
estimate the neutrino resonance conditions. The solutions of the
equation $\frac{d D_{(n)}}{d A} = 0$ will give the the turning value of
the discriminant $D_{(n)}$. Although, analyses of the
derivative of the discriminant $D_{(n)}$ for 3- and 4-neutrino models
are complicated, it is expected that the solution of the equation 
$\frac{d D_{(n)}}{d A} = 0$, that is a resonance condition, is given as
an equation for $A$ ,$E$, and the function of some parameters: 
\begin{equation}
 A E = f(\Delta m_{ab}^2, \theta_{ab}, \Delta_{ab}).
  \label{eq:solution-derivative}
\end{equation}
If the value in the right hand side of the
(\ref{eq:solution-derivative}) is fixed, e.g, as in
(\ref{eq:DeltaMsolar}), (\ref{eq:DeltaMatm}), (\ref{eq:DeltaMlsnd}) and
(\ref{eq:MixingAngles}), we can conclude that $E$ and
$A$ are in inverse proportion to each other. 

\subsection{Normalized discriminant}
In this analysis, the discriminant of the characteristic equation for
the Hamiltonian in matter has a small value. The value of the discriminant in
vacuum is given by setting the matter density $A$ to $0$.
Here we assume that the electron number density $N_e$ is equal to
the neutral particle number density $N_n$ : $N_e = N_n$ which means
$A_e=A$ and $A_n=\frac{1}{2}A$ in (\ref{eq:defA-e}) and (\ref{eq:defA-n}).

The effective energy differences
$|\lambda_a - \lambda_b|$ $\quad$ $(a,b=1,2,3,4, a \neq b)$, 
which are the differences between eigenvalues $\lambda_a$ of the
Hamiltonian in matter, are functions of the matter density $A$. 
If we set $A=0$, then the effective energies $\lambda_a$ can be regard
as the eigenvalues $E_a$ of the Hamiltonian in vacuum:
\begin{equation}
 |\lambda_a - \lambda_b|=|E_a - E_b| \quad \mathrm{for} \, A=0.
\end{equation}

In a 2-neutrino ($\nu_e$, $\nu_{\mu}$) model, from
(\ref{eq:differrenceEab}), we get a rough estimation for the
discriminant $D_{(2)}$:
\begin{eqnarray}
 D_{(2)}\big|_{A=0} &=& {(\lambda_2 - \lambda_1)^2}\big|_{A=0}
  = (E_2-E_1)^2 
  \simeq \left[
     \frac{\Delta {m_{21}}^2}{2E}
     \right]^2
  \nonumber \\
 &=& 
  \left[
   \frac{\Delta m_{\mathrm{solar}}^2}{2E}
  \right]^2
  \sim 10^{-29} [\mathrm{eV}^2] \quad 
  (\mathrm{for} \  E=10^{10} \mathrm{eV}).
  \nonumber\\
\end{eqnarray}
It may be easy to expect that, in the 3- and 4-neutrino models, 
the discriminant $D_{(3)}$, $D_{(4)}$ in vacuum have very small value, since
the discriminant are defined by the product
(\ref{eq:discriminant-def}). Indeed,
$D_{(3)} \sim 10^{-82} [\mathrm{eV}^6]$ and 
$D_{(4)} \sim 10^{-142} [\mathrm{eV}^{12}]$ for $A_e=A_n=0$, e.g., in
vacuum, by our rough calculations. Here we set $E=10\,\mathrm{GeV}$.

In order to concentrate to the argument about the behavior for $D_{(n)}$
as the function of $A$, we define the normalized discriminant $D_n$
which is expressed as the value of the $D_{n}(A)$ divided by that for
$A=0$:
\begin{equation}
 D_n = \frac{D_{(n)}(A)}{D_{(n)}{(A=0)}}.
\end{equation}

Hereafter, $D_4$ means the normalized discriminant for the 
$4 \times 4$ matrix which is the operator of the 4-neutrino
($\nu_e, \nu_{\mu}, \nu_{\tau}, \nu_s$) model. Moreover, $D_3$
specializes the normalized discriminant for the $3 \times 3$ matrix of
Hamiltonian, in which we deal with the three active neutrinos: 
($\nu_e, \nu_{\mu},\nu_{\tau}$).
In order to compare the 3- and 4-neutrino models with the usual neutrino
phenomenology, we consider the two neutrinos 
($\nu_e, \nu_{\mu}$) for ${D_2}$.

\subsection{Graphical expression of resonance conditions}
As an illustration of the resonance phenomena, the normalized
discriminants $D_n$ $(n=2,3,4)$ are plotted as a function of the
matter density $A$ in Fig. \ref{fig:DnE010G}. Here we set the average
$E$ of the neutrino energies to $10\,\mathrm{GeV}$, which is a value
of neutrino energy used in previous analyses \cite{Ohlsson00,Kamo03}.  

From Fig. \ref{fig:DnE010G}, the following results can be derived. If
the mixing angles $\theta_{13}$ and $\theta_{14}$ are small, 
the third- and fourth-neutrino effects appear as a resonance. The sharp
drop in the Fig. \ref{fig:DnE010G} is a good reflection of the neutrino
resonance. Indeed, the surviving probabilities $P_{ee}$, which are
discussed in \cite{Ohlsson00} and \cite{Kamo03}, for neutrino
oscillations in matter have sharp drops. The smaller a mixing angle
reduces, the sharper the drops for the discriminant $D_{(n)}$ and the
surviving probabilities become.  

One effect for the third neutrino $\nu_{\tau}$ appears beyond 
$A \sim 10^{-13} \, \mathrm{eV}$. The other effect for the fourth
neutrino $\nu_s$ also occurs beyond $A \sim 10^{-10} \, \mathrm{eV}$. 
It means that the matter density to cause the resonance for 
$\nu_e \leftrightarrow \nu_{s}$ or $\nu_e \leftrightarrow \nu_{\tau}$
varies with the squared mass differences:
\begin{equation}
 \frac{10^{-10}}{10^{-13}} = 10^{3} \sim
  \frac{\Delta m_{\mathrm{LSND}}^2}{\Delta m_{\mathrm{atm}}^2}.
\end{equation}

The above results are argued for $E=10\,\mathrm{GeV}$. The normalized
discriminant for $E=10\,\mathrm{MeV}$ is shown in Fig. \ref{fig:DnE010M}. 
The lines in Fig. \ref{fig:DnE010G} and Fig. \ref{fig:DnE010M} have the
similar shapes except for the value of the matter density $A$.
The difference of two results occurs due to the difference of the
average $E$ of neutrino energies. For instance, one resonance for 
$\nu_e \leftrightarrow \nu_{\tau}$ in
Fig. \ref{fig:DnE010G} ($E=10\,\mathrm{GeV}$) occurs for 
$A \sim 10^{-13} \, \mathrm{eV}$, the other resonance for 
$\nu_e \leftrightarrow \nu_{\tau}$ in Fig. \ref{fig:DnE010M} 
($E=10\,\mathrm{MeV}$) occurs at $A \sim 10^{-10} \, \mathrm{eV}$. 
Therefore the average $E$ of the neutrino
energies and the matter density $A$ are in inverse
proportion to each other as shown in
(\ref{eq:ResonanceConditon2NuGeneral}), i.e., 
\begin{equation}
 E A = 
  \frac{\Delta m_{\mathrm{atm}}^2}{2}\cos2\theta_{13} 
  \sim 10^{-3} [\mathrm{eV}^2].
  \label{eq:ResonanceConditionsNu13}
\end{equation}
Note that we set the matter density to $A_e=A$, $A_n=(1/2)A$.
In the case of the resonance $\nu_e \leftrightarrow \nu_{s}$,
\begin{equation}
 E A = \Delta m_{\mathrm{LSND}}^2 \cos2\theta_{14} 
  \sim 1 [\mathrm{eV}^2].
  \label{eq:ResonanceConditionsNu14}
\end{equation}

\section{Discussion}
\label{chap:Discussion}
\begin{figure}[htb]
 \begin{center}
  \includegraphics[height=6cm,clip]{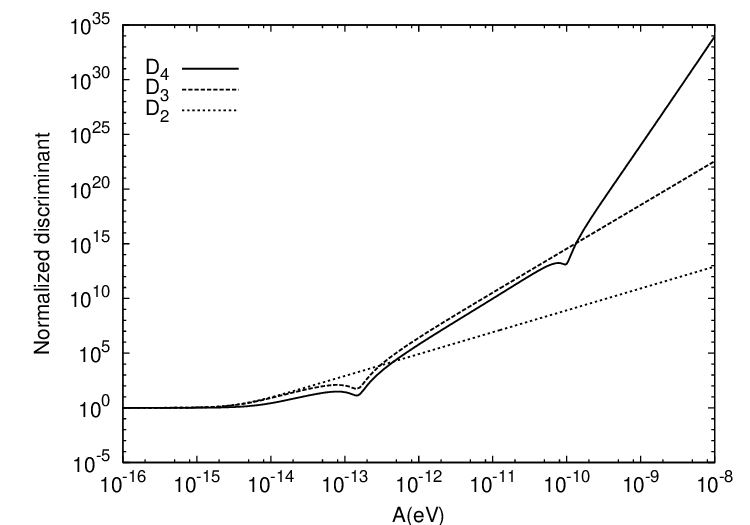}
 \end{center}
  \caption{Normalized discriminants $D_n$ $(n=2,3,4)$ as a function of
 the matter density $A$ for a $(3+1)_1$-scheme, where
 $\theta_{12} = \frac{\pi}{4}$, $\theta_{23} = \frac{\pi}{4}$,
 $\theta_{13} = \frac{5}{180}\pi$, $\theta_{14} = \frac{3}{180}\pi$,
 $\theta_{24} = 0$, $\theta_{34} = 0$,
 $\Delta_{13} = \Delta_{24} = \Delta_{34} = 0$,
 $\Delta m_{21}^2 \simeq 7 \times 10^{-5} \,\mathrm{eV^2}$,
 $\Delta m_{32}^2 \simeq 3 \times 10^{-3} \,\mathrm{eV^2}$, 
 $\Delta m_{41}^2 \simeq 1 \,\mathrm{eV^2}$,
 $E=10$ GeV. The solid, broken and dotted lines are normalized
 discriminants $D_4$, $D_3$ and $D_2$, respectively.}
  \label{fig:DnE010G}
\end{figure}

\begin{figure}[htb]
 \begin{center}
  \includegraphics[height=6cm,clip]{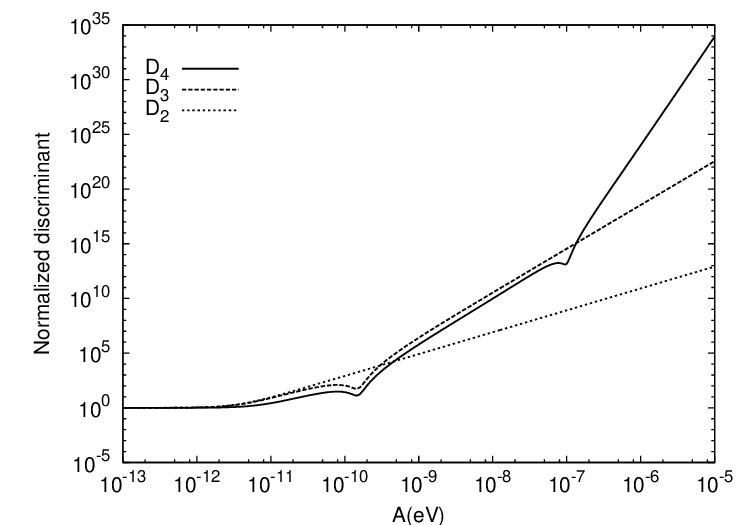}
 \end{center}
  \caption{Normalized discriminants $D_n$ $(n=2,3,4)$ as a function of
 the matter density $A$ for a $(3+1)_1$-scheme, where
 $\theta_{12} = \frac{\pi}{4}$, $\theta_{23} = \frac{\pi}{4}$,
 $\theta_{13} = \frac{5}{180}\pi$, $\theta_{14} = \frac{3}{180}\pi$,
 $\theta_{24} = 0$, $\theta_{34} = 0$,
 $\Delta_{13} = \Delta_{24} = \Delta_{34} = 0$,
 $\Delta m_{21}^2 \simeq 7 \times 10^{-5} \,\mathrm{eV^2}$,
 $\Delta m_{32}^2 \simeq 3 \times 10^{-3} \,\mathrm{eV^2}$, 
 $\Delta m_{41}^2 \simeq 1 \,\mathrm{eV^2}$,
 $E=10$ MeV. The solid, broken and dotted lines are normalized
 discriminants $D_4$, $D_3$ and $D_2$, respectively.}
  \label{fig:DnE010M}
\end{figure}

The main result of our analysis is derivation of the discriminant of the
characteristic equation for the Hamiltonian in matter, which is
discussed in flavor basis. The concrete form of the discriminant for 2-,
3-, and 4-neutrino models is given in (\ref{eq:discriminant2}), 
(\ref{eq:discriminant3}), and (\ref{eq:discriminant4}), respectively.
The effective Hamiltonian in the 2-neutrino model is also calculated
concretely. The discriminant made from the effective Hamiltonian is a
function of the matter density $A$ and the average $E$ of the neutrino
energies. The calculations for 3- and 4-neutrino models are
straightforward but the forms of the discriminants are complicated.
One of the  numerical expression of the discriminant is given in
(\ref{eq:D3num}) and (\ref{eq:D4num}). 
The graphical expression of the discriminant are also shown in
Fig.\ref{fig:DnE010G} and Fig.\ref{fig:DnE010M}.

One of the resonance conditions for the neutrino oscillation is
described in (\ref{eq:ResonanceConditon2NuGeneral}). It shows that $E$
and $A$ are in inverse proportion to each other, as confirmed in 
Fig. \ref{fig:DnE010G} and Fig. \ref{fig:DnE010M}. That is, the lower
the neutrino energy reduces the higher
the density to observe the resonance between the neutrinos becomes.
The concrete form of the resonance condition for 2-neutrino model is
shown in (\ref{eq:ResonanceConditionsNu13}) and
(\ref{eq:ResonanceConditionsNu14}).
From (\ref{eq:D3num}) and (\ref{eq:D4num}), the discriminant $D_{(n)}$
is a smooth function of $A$. The derivative $\frac{d D_{(n)}}{d A}$ of
$D_{(n)}$ may be useful to estimate the neutrino resonance
conditions. Although, analyses of the derivative of the discriminant
$D_{(n)}$ for 3- and 4-neutrino models are complicated, we can expect
that the neutrino resonance condition will be given as
(\ref{eq:solution-derivative}). 

\begin{figure}[htb]
 \begin{center}
  \includegraphics[height=6cm,clip]{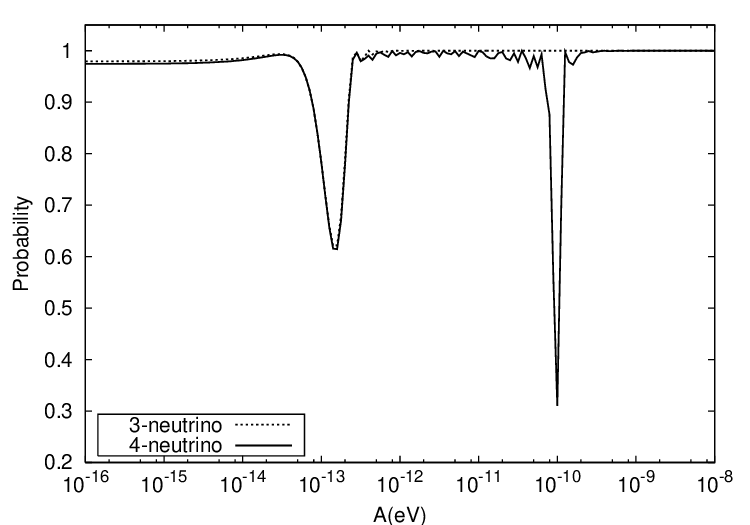}
 \end{center}
  \caption{The surviving probability $P_{ee}$ of the electron neutrino
 transition as a function of the matter density $A$ for the
 $(3+1)_1$-scheme, where $\theta_{12} = \frac{\pi}{4}$, 
 $\theta_{23} = \frac{\pi}{4}$,
 $\theta_{13} = \frac{5}{180}\pi$, $\theta_{14} = \epsilon$,
 $\theta_{24} = 0$, $\theta_{34} = 0$,
 $\Delta_{13} = \Delta_{24} = \Delta_{34} = 0$,
 $\Delta m_{21}^2 \simeq 7 \times 10^{-5} \,\mathrm{eV^2}$,
 $\Delta m_{32}^2 \simeq 3 \times 10^{-3} \,\mathrm{eV^2}$, 
 $\Delta m_{41}^2 \simeq 1 \,\mathrm{eV^2}$,
 $E=10\, \mathrm{GeV}$ and $L=10000 \, \mathrm{km}$.
 The solid and broken lines show the transition probabilities for
 $\epsilon = \frac{3}{180}\pi, 0$, respectively, which correspond to the
 4-neutrino
 and the 3-neutrino cases.}
  \label{fig:PexE010GL10000km}
\end{figure}

The conditions that the discriminant $D_{(n)}$ has a local minimum value
give the resonance conditions for neutrino oscillation. In our analysis,
it is expected that the resonance of $\nu_e \leftrightarrow \nu_{\tau}$
and $\nu_e \leftrightarrow \nu_{s}$ occur at 
$A=10^{-13}\, \mathrm{eV}$ and $A=10^{-10}\, \mathrm{eV}$, respectively,
for $E=10 \, \mathrm{GeV}$. This results can be confirmed by the
previous calculations \cite{Ohlsson00,Kamo03} in 2-, 3- and 4-neutrino
models, respectively. 
For instance, the surviving probabilities $P_{ee}$ for 4-neutrino
oscillation is given as
\begin{equation}
 P_{ee}
  =
  1 -4 
  \sum_{\scriptstyle a=1}^{3}
  \sum_{\scriptstyle b=a+1}^4
  {B_a B_b} 
  \sin ^2 \left[
	   \frac{L}{2} \cdot 
	   \frac{{\Delta{m}_{ab}}^2}{2{E}}
	  \right],
  \label{eq:transitionNueNue}
\end{equation}
where $B_a$ is a factor determined by the mixing matrix and
the characteristic equation of the matter Hamiltonian in mass basis%
\cite{Kamo03}.
The factor $L$ is a baseline length which stands for the neutrino flying
distance. Fig. \ref{fig:PexE010GL10000km} is a result of the
recalculation of the transition probabilities for new conditions which
we have used in this paper. One can find that the value of the matter
density $A$ that the resonance occurs in Fig. \ref{fig:DnE010G}
coincides with the value of $A$ that the sharp drop of the transition
probabilities occurs in Fig.\ref{fig:PexE010GL10000km}.  

Is the condition to observe the neutrino resonance realistic?
The electron matter density $A$ is $1 \times 10^{-11} \, \mathrm{eV}$
at the center of the sun. The average of the electron matter density in
earth is about $5 \times 10^{-13} \, \mathrm{eV}$ \cite{Kamo03}. 
In the case of $E=10 \, \mathrm{GeV}$, it may be possible to observe the
resonance between $\nu_e$ and $\nu_{\tau}$ since the condition is 
$A \sim 10^{-13} \, \mathrm{eV}$ which corresponds to the density of the
earth. But in other analysis it may be difficult to observe the
resonance.

One may consider that the resonance will occur in the high energy and
low matter density, which is restricted by
(\ref{eq:ResonanceConditon2NuGeneral}). But there is a limit of the
neutrino energy to observe the neutrino oscillation. From 
(\ref{eq:transitionNueNue}), the baseline length $L$ to observe the
neutrino oscillation and the average $E$ of the neutrino energies are
proportional to each other. For instance, the condition to observe a
neutrino transition 
$\nu_e \to \nu_{s}$ for $E=10 \,\mathrm{GeV}$ and $E=10 \,\mathrm{MeV}$
is $L>25000 \, \mathrm{m}$ and $L>25 \, \mathrm{m}$, respectively, where
we assume the condition 
\begin{equation}
 \frac{L}{\Lambda/4} = \frac{{\Delta m_{ab}}^2 \cdot L}{2\pi E} > 1.
\end{equation}
Then, we have considered that the neutrino oscillation can be observed when
the neutrino flying distance is more than $\Lambda/4$, which $\Lambda$
corresponds to the wavelength in (\ref{eq:transitionNueNue}).

The 3-neutrino model with a small and bi-maximal mixings and the
4-neutrino model will be tested more closely by the upcoming
experiments. In this paper, the resonance conditions for 2-, 3-, and
4-neutrino mixings are given without a preconception. Our result of the
resonance conditions, which are given by using the discriminant, does
not deny the possibility of the 4-neutrino oscillation, but the
observation of the resonance is not realistic.

\appendix
%\newpage

%\newpage
%%% Fig. massscheme

\newpage
%%% Fig.normalized discriminant for E=10GeV

%%% Fig.normalized discriminant for E=10MeV

%%% Fig. e-neutrino surviving probabilities

\end{document}